\documentclass[aps,prx,twocolumn,superscriptaddress,showpacs]{revtex4-1}
\usepackage{amsmath,amssymb,graphics,epsfig,epstopdf,color,verbatim,ulem,braket,tabularx}
\usepackage{multirow}
\usepackage[colorlinks,linkcolor=blue,citecolor=blue,urlcolor=blue,bookmarks=false]{hyperref}
\usepackage{listings}

\begin{document}

\title{Symmetry Enforced Chiral Hinge States and Surface Quantum Anomalous Hall Effect in Magnetic Axion Insulator $\text{Bi}_{2-x}\text{Sm}_x\text{Se}_3$}

\author{Changming Yue}
\affiliation{Beijing National Laboratory for Condensed Matter Physics, and Institute of Physics, Chinese Academy of Sciences, Beijing 100190, China}
\affiliation{University of Chinese Academy of Sciences, Beijing 100049, China}
\author{Yuanfeng Xu}
\affiliation{Beijing National Laboratory for Condensed Matter Physics, and Institute of Physics, Chinese Academy of Sciences, Beijing 100190, China}
\affiliation{University of Chinese Academy of Sciences, Beijing 100049, China}
\author{Zhida Song}
\affiliation{Beijing National Laboratory for Condensed Matter Physics, and Institute of Physics, Chinese Academy of Sciences, Beijing 100190, China}
\affiliation{University of Chinese Academy of Sciences, Beijing 100049, China}
\author{Hongming Weng}
\affiliation{Beijing National Laboratory for Condensed Matter Physics, and Institute of Physics, Chinese Academy of Sciences, Beijing 100190, China}
\affiliation{Collaborative Innovation Center of Quantum Matter, Beijing, China}
\author{Yuan-Ming Lu}
\affiliation{Department of Physics, The Ohio State University, Columbus, OH 43210, USA}
\author{Chen Fang}
\affiliation{Beijing National Laboratory for Condensed Matter Physics, and Institute of Physics, Chinese Academy of Sciences, Beijing 100190, China}
\author{Xi Dai}
\email{daix@ust.hk}
\affiliation{Department of Physics, Hong Kong University of Science and technology, Clear Water Bay, Kowloon, Hong Kong}
\affiliation{Beijing National Laboratory for Condensed Matter Physics, and Institute of Physics, Chinese Academy of Sciences, Beijing 100190, China}

\begin{abstract}
A universal mechanism to generate chiral hinge states in the ferromagnetic axion insulator phase is proposed, which leads to an exotic transport phenomena, the quantum anomalous Hall effect (QAHE) on some particular surfaces determined by both the crystalline symmetry and the magnetization direction. A realistic material system Sm doped $\text{Bi}_2\text{Se}_3$ is then proposed to realize such exotic hinge states by combing the first principle calculations and the Green's function techniques. A physically accessible way to manipulate the surface QAHE is also proposed, which makes it very different from the QAHE in ordinary 2D systems.
\end{abstract}

\maketitle

\newcommand{\alert}[1]{{\color{red}{#1}}}

The bulk-boundary correspondence \cite{BBC_TopoOrig_Yasuhiro} is one of the most important physical consequences of the topological matter. In most of the cases, the bulk-boundary correspondence refers to the existing of guaranteed gapless quasi-particle excitations on the ($d-1$)-dimensional boundary (with $d$ being the dimension of the system), given that the symmetries required to protect such a topological state are still preserved on the boundary \cite{Hatsugai1993,Kane2005,Fu2011,Hsieh2012,Wang2016}. Very recently a new type of  bulk-boundary correspondence was proposed for a special class of topological materials called second order topological insulators (SOTIs) \cite{PhysRevB.96.245115, PietW_Reflection_SOTI, PhysRevB.97.155305, Titus_HOTI, zhida_prl_dm2_edge, Fang2017, Bernevig_MulIns,Bi_HOTI}, where the corresponding topological quasi-particle states appear in the $(d-2)$ instead of $(d-1)$-dimensional boundaries. In particular, non-trivial corner states (0D) will appear at the corners of a two dimensional SOTI and helical/chiral hinge states (1D) will appear at some particular hinges of a three dimensional SOTI.

Similar to the ordinary TI \cite{Kane2005,RevModPhys.82.3045, RevModPhys.83.1057,annurev_TCI_TSC_LiangFu}, the general definition of SOTI can be expressed as the band insulators that can not be smoothly deformed to the atomic insulators \cite{Soluyanov2011,Po2017,Bradlyn2017} without symmetry breaking or closing the bulk energy gap. Unlike topological insulators where each surface is gapless, a generic ($d-1$)-dimensional surface of $d$-dimensional SOTI is gapped, but on the entire boundary of a sample, there must be ($d-2$)-dimensional domain walls between surfaces having opposite masses, at which are located gapless topological modes. For the 2D SOTI, additional chiral or charge conjugation symmetry is required to protect the corner states \cite{PietW_Reflection_SOTI}, which is difficult to realize in materials. While for 3D SOTI, the chiral or helical hinge states can be enforced by some bulk crystalline topological invariants, i.e. the mirror Chern number (MCN) $C_m$, but protected by more general symmetries\cite{zhida_prl_dm2_edge,Fang2017}. For example, for the helical hinge states, as first introduced in Ref. \cite{Hsieh2012,Titus_HOTI}, the existence of non-trivial hinge states can be enforced by the non-zero mirror Chern number $C_m=2N$ defined on some particular mirror invariant planes with $N$ being an odd integer. The non-trivial helical states will persist even when the mirror symmetry is no longer present and hence MCN cannot be defined. As long as both the bulk and surfaces around that particular hinge are all fully gapped, the only symmetry requirement to protect the non-trivial helical mode in such cases is the time reversal symmetry. In Ref. \cite{Titus_HOTI}, SnTe with strain along the (100) direction is proposed to be the first realistic material which supports the non-trivial helical states on the hinge formed between (100) and (010) surfaces.

\begin{figure*}[htp]
\includegraphics[clip,width=0.8\paperwidth,angle=0]{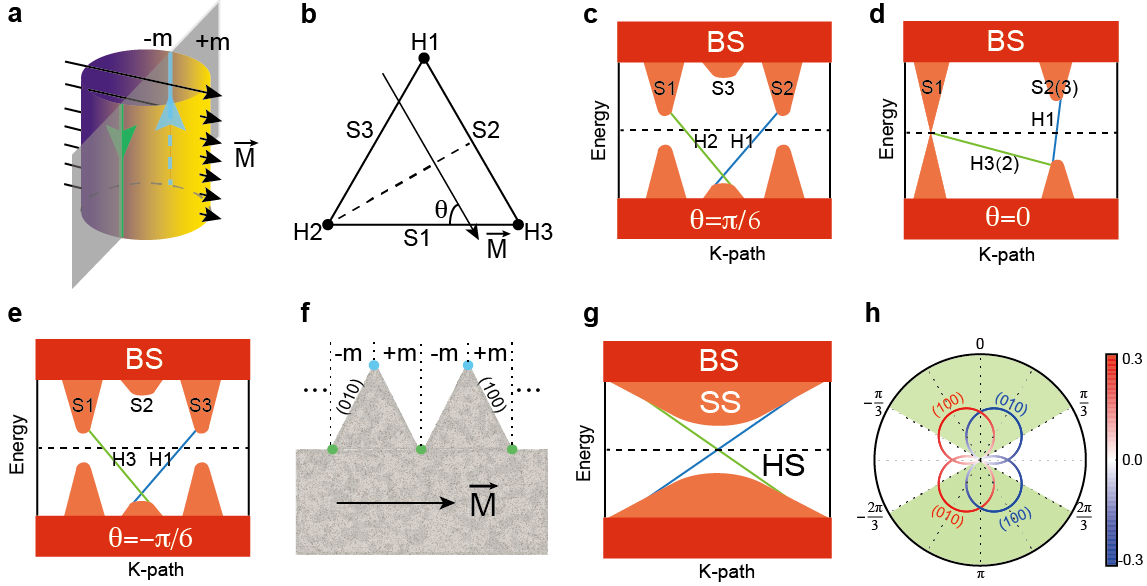}
\caption{Schematic demonstration of the emergence of hinge states.
(a) Chiral hinge states emerges on the domain wall of the surface states with magnetic polarization applied perpendicular to the mirror plane. (b) 3-D TI with $C_{3v}$ symmetry, and (c) to (e) schematically illustrate the evolution of its surface states and hinge states with the magnetic polarization applied along different directions. (f) The proposed configuration of the magnetically doped $\text{Bi}_{2}\text{Se}_{3}$  sample that can be adopted to detect the chiral hinge states. (g) The schematic plot of the energy dispersion of the corresponding hinge states in (f). (h)The ratio $m/|\vec{\text{M}}|$ for (010) and (100) surface states as a function of the direction of magnetization, $\theta$. The green shaded area labels the range of $\theta$ where hinges state are formed. For (a) and (c) to (g), blue and green lines denote two hinge states with opposite chirality.}\label{schematic}
\end{figure*}

Compared to helical hinge states, chiral hinge states are even more stable, which requires no symmetry to protect them. Similar to the chiral edge states in the 2D quantum Hall systems \cite{Klitzing1980,Hatsugai1993}, the chiral hinge states associated with a specific surface will lead to surface quantum anomalous Hall effect (QAHE) \cite{Mong2010,Nomura2011}, which is the QAHE \cite{Haldane1988,Yu2010,Chang2013} on a completely three dimensional object and has never been observed in any condensed matter systems before. In the present letter, we will propose that both the chiral hinge states and hence the surface QAHE can be realized in ferromagnetic axion insulators. After that we further propose a realistic material system, Sm doped $\text{Bi}_2\text{Se}_3$ single crystal, to realize such an axion insulator and SOTI with chiral hinge states. The high quality single crystal of Sm doped $\text{Bi}_2\text{Se}_3$ has been already obtained in Ref. \cite{Sm_dopeBi2Se3} with very low carrier density and Curie temperature as high as 52K. The easy axis of the magnetization has been confirmed experimentally to be within the $ab$ plane. From our DFT calculations, the effective exchange field acting on the low-energy bands is around 20 meV, which is much smaller than the semiconductor gap in $\text{Bi}_2\text{Se}_3$, thus keeping the system within the axion insulator phase. The presence of the chiral hinge states can be illustrated schematically in Fig.~\ref{schematic}(b). Since the crystal structure of $\text{Bi}_2\text{Se}_3$ contains three vertical mirror planes, one of them can survive the ferromagnetic order by putting the magnetization direction perpendicular to that mirror plane as shown in Fig.~\ref{schematic}(b).  The original Dirac surface states on different  sides of the mirror plane (S1 and S3) will acquire finite masses, which have opposite signs forced by the mirror symmetry as illustrated in Fig.~\ref{schematic}(b). Therefore on hinge H2 between surface S1 and S3 in  Fig.~\ref{schematic} (b), a domain wall between massive Dirac surface states with different mass signs is enforced by the mirror symmetry,  leading to 1D chiral states on the corresponding hinge. Since the 1D chiral mode is stable against any weak perturbations, when the Zeeman field rotates away from the symmetric position and the system no longer has a mirror symmetry, the chiral hinge states cannot disappear immediately. However its location can be modified or even moved from one hinge to another once the gap on surface S1 or S3 is closed and reopen, as illustrated schematically in Fig. ~\ref{schematic} (c) to (e).

The above argument can be made more general. In quantum electrodynamics, Lorentz invariance allows, in addition to the Maxwell term, a topological term of the form $\frac{\theta}{4\pi}\mathbf{E}\cdot\mathbf{B}$, where the coupling strength $\theta$ is called the axion field\cite{Wilczek1987}. A static $\theta$ is quantized to $0$ or $\pi$ in the presence of time-reversal symmetry \cite{Qi2008}, and $\theta=0$ and $\pi$ for trivial and topological insulators respectively. It was then realized that in the absence of time-reversal, spatial symmetries can also quantize the value of $\theta$, inducing a natural generalization of topological insulators\cite{Essin2009}. In fact, as long as the spatial symmetry is \textit{improper}, i.e. it flips the orientation of the space, a uniform $\theta$ can only take $0$ or $\pi$. It is usually thought that at the boundary of a region with $\theta=\pi$, there are gapless 2D surface states, which are nothing but the single Dirac fermion in the case of topological insulators. However, if the symmetry quantizing $\theta$ is a spatial symmetry, which is the case for the FM axion insulators, the 2D surface states only exist if the interface preserves the symmetry, and on a surface where it is broken, a mass gap generically exists. This leads to another interesting possibility: the mass is enforced to change signs on the entire boundary of the topological state, creating domain walls of mass gaps\cite{PhysRevLett.110.046404}, because either the mirror reflection or the inversion symmetry flips the sign of the mass terms for the surface Dirac Hamiltonian. Along any one of these domain walls, there are 1D chiral modes, i.e. the $(d-2)$-dimensional topological edge states, the recently discovered boundary manifestation of the nontrivial topology in the bulk in the absence of gapless surface states. In Fig.~\ref{schematic}(a), we illustrate the 1D chiral modes on the surface of 3D axion insulators protected by the improper symmetry, mirror reflection in this case.

To calculate the hinges states, we adopt an approximation where the low-energy physics of Sm doped $\text{Bi}_2\text{Se}_3$ can
be modeled by the tight binding Hamiltonian for the pure $\text{Bi}_2\text{Se}_3$ together with a magnetic exchange field acting
on the p-orbitals of both the Bi and Se atoms.  Moreover, we generalize the recursive Green's function method\cite{Green_recur1985},
which is widely used for the spectral functions of the surface or interface states, to calculate the hinge states for the SOTI (details given in appendix B). As shown in Fig.~\ref{hinge-recursive}(c),
two hinges (labeled as Left and Right respectively) are generated by joining two semi-infinite slabs along different directions,
namely $[1,0,0]$ and $[0,1,0]$. The hinge area can then be viewed as the left and right ends of the interface region between
the two slabs oriented along different directions, with the cross section perpendicular to the c-axis.
The projected spectral functions for the two hinge regions can then be obtained following the
standard procedures of the recursive Green's function method \cite{Green_recur1985}.
\begin{figure}[htp]
\includegraphics[clip,width=3.4in,angle=0]{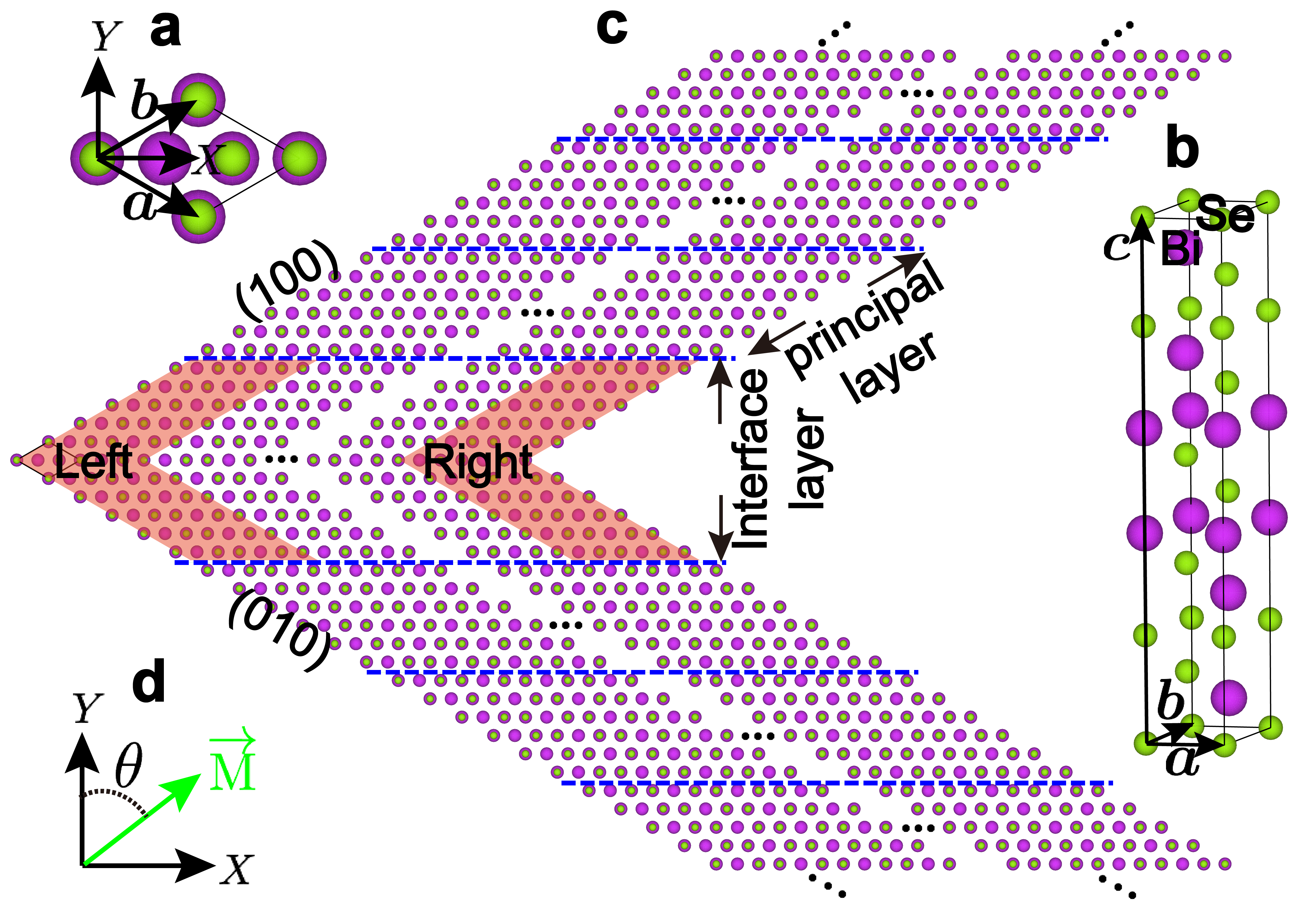}
\caption{The structure proposed for the calculation of hinge states. (a), (b) Crystal structure of $\text{Bi}_2\text{Se}_3$. $X$, $Y$ and $Z$
form the global right-handed coordinate system.
(c) The bi-semi-infinite open boundary geometry in which the (010) and (100) surfaces
meet at \emph{left} and \emph{right} hinges parallel to $\boldsymbol{c}$ direction. The structure
is semi-infinite along $\boldsymbol{a}$ and $\boldsymbol{b}$ direction, periodic along $\boldsymbol{c}$
direction, finite along $X$ direction. The length in $X$ direction is $20$ lattice constants.
\emph{Principal layer} (PL) is a group of atomic layers that is large enough
such that only adjacent PLs interact. \emph{Interface layer} is a group of atomic layers to connect
two semi-infinite parts. Generally, interface layer is composed of the first PL along $\boldsymbol{a}$ direction
combined with the first PL along $\boldsymbol{b}$ direction. (d) The easy direction of the magnetization $\boldsymbol{M}$ lies in the $ab$ plane,
with $\theta$ the angle between $\boldsymbol{M}$ and $y$ axis.
}\label{hinge-recursive}
\end{figure}
As already confirmed experimentally \cite{Sm_dopeBi2Se3}, in $\text{Sm}_x\text{Bi}_{2-x}\text{Se}_3$ the magnetic moments on  Sm  ions  ordered ferromagnetically
under $T_c\approx 52$K. The easy axis lies within the $ab$ plane and can be easily tuned by a small external magnetic field.
The crystal structure of $\text{Bi}_2\text{Se}_3$ contains three vertical mirror planes and will survive the ferromagnetic order if the magnetization
$M$ is along the $[-1,1,0]$, $[0,-1,0]$ and $[1,0,0]$ directions respectively, (corresponding to $\theta= 0, 2\pi/3$ and $4\pi/3$ in Fig.~\ref{schematic}(b) respectively).
Then as we discussed above, the existence of the
mirror symmetry in the axion insulator will force the mass terms on the different sides of the mirror symmetric hinge to be opposite in sign,
which guarantees the existence of the chiral hinge states centered at that particular hinge. Interestingly, since the
chiral hinge states are topologically stable, the breaking of the corresponding mirror symmetry, i.e. by rotating the magnetization $M$
away from the particular angle mentioned above, the chiral hinge states won't disappear immediately. In fact it will disappear only when the
surface gap closes on either of the two nearby surfaces. Otherwise, as long as the gap still exists on both surfaces near the hinge, the domain wall feature still remains and the only effect of the mirror symmetry breaking is to modify
the wave function of the hinge state to be asymmetric about the mirror plane.

The local spectral functions at both the left and right hinges can be obtained by projecting the imaginary part of the Green's
function to the corresponding hinge area, which can be expressed as $\rho_{L/R}(k_z,\omega)={-1\over\pi}\sum_{i\in L/R} \text{Im}G_{ii}(k_z,\omega)$,
where the hinge area $L$ and $R$ are illustrated by the orange color shaded block
\emph{Left} and \emph{Right} in Fig.~\ref{hinge-recursive}(c), respectively. Then the hinge spectral
functions can be obtained by applying the recursive Green's function method introduced above. Alternatively the existence of chiral hinge modes can
be inferred from checking the mass terms on the nearby surfaces, whose Hamiltonian can be written as
\begin{equation}
H_{SF} = {\vec v}_x\cdot {\vec \sigma}k_x
+ {\vec v}_y\cdot {\vec \sigma}k_y + {\vec g}_x\cdot {\vec \sigma}M_x + {\vec g}_y\cdot {\vec \sigma}M_y +{\vec g}_z\cdot {\vec \sigma}M_z,
\end{equation}
where $x$, $y$, $z$ form the local right handed coordinate system and ${\vec v}_i$, ${\vec g}_i$ are vectors defined in pseudo-spin space.
Both the velocity and g-factor vectors for the (100) and (010) surfaces of Sm$_x$Bi$_{2-x}$Se$_3$ can be obtained by the corresponding
surface calculations based on the effective tight binding Hamiltonian with their values listed in Table.~\ref{vgtable}.
\begin{figure*}[htp]
\includegraphics[clip,width=0.8\paperwidth,angle=0]{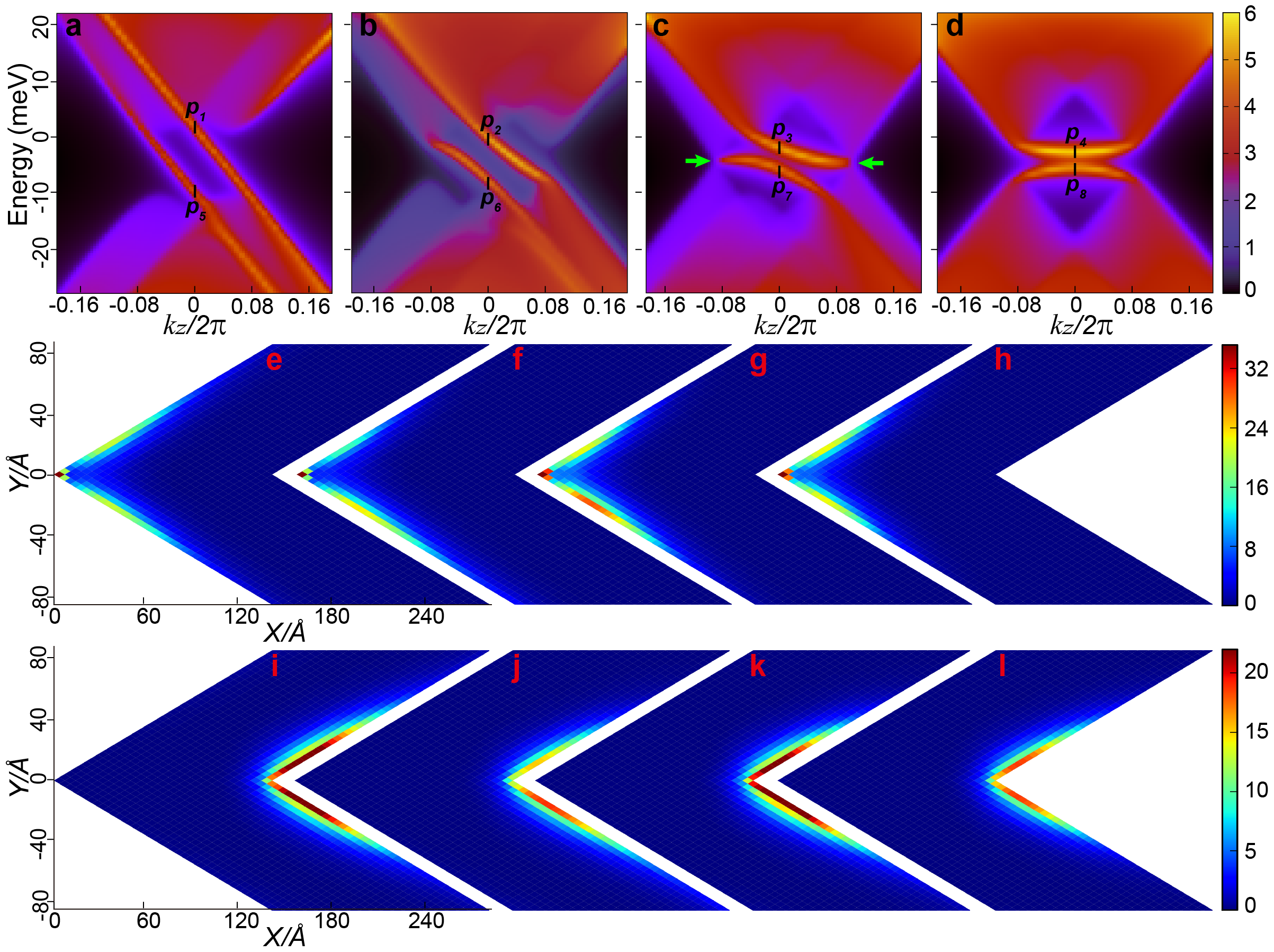}
\caption{The projected spectral function (a-d) on the interface layer and their spatial distributions (e-l). From (a) to (d), $\boldsymbol{M}$ is rotated from
$\theta=0$ to $\theta=\pi/6$, $\theta=\pi/3$ and $\theta=\pi/2$, respectively. Points $\{p_1,\cdots,p_8\}$, with $k_z=0$, are labelled to show spatial distribution of 
these points in (e) to (l). From (e) to (h), we plot spatial distribution of spectral functions at $p_1$ to $p_4$ labelled in (a) to (d), showing spatially localized states distributed around the
left hinge. From (i) to (l), we plot spatial distribution of spectral functions at $p_5$ to $p_8$ labelled in (a) to (d), showing spatially localized states distributed around the
right hinge.}\label{hinge_spectra_distr}
\end{figure*}
\begin{table}
\centering
\caption{Local right-handed coordinate system ${\boldsymbol{e}_{x,y,z}}$, velocity $v$ and g-factor $g$ of (010) and (100) surfaces, respectively.
Entries of local $\boldsymbol{e}_{x,y,z}$ are given in terms of unit vectors of global coordinate system
$x,y,z$ shown in Fig.~\ref{hinge-recursive}-(a). Unit of $v$ is eV$\cdot\AA$ while unit of $g$ is 1/eV.
} \label{vgtable}
\begin{ruledtabular}
\begin{tabular}{ccccccc}
&\multicolumn{3}{c}{(010)}&\multicolumn{3}{c}{(100)} \\
\hline
$\boldsymbol{e}_{x}$ & $\frac{\sqrt{3}}{2}$ & $-\frac{1}{2}$ & 0 & $-\frac{\sqrt{3}}{2}$ & $-\frac{1}{2}$ & 0 \\
$\boldsymbol{e}_{y}$ & 0 & 0 & 1 & 0 & 0 & 1 \\
$\boldsymbol{e}_{z}$ & $-\frac{1}{2}$ & $-\frac{\sqrt{3}}{2}$ & 0 & $-\frac{1}{2}$ & $\frac{\sqrt{3}}{2}$ & 0 \\
$\vec{v}_{x}$ & 0 & 0.4316 & 0.4951 & 0 & 0.4316 & 0.4951 \\
$\vec{v}_{y}$ & -0.7649 & 0 & 0 & -0.7649 & 0 & 0 \\
$\vec{g}_{x}$ & 0.7782 & 0 & 0 & 0.7782 & 0 & 0 \\
$\vec{g}_{y}$ & 0 & 0.5354 & -0.0957 & 0 & 0.5354 & -0.0957 \\
$\vec{g}_{z}$ & 0 & 0 & 0.4778 & 0 & 0 & 0.4778 \\
\end{tabular}
\end{ruledtabular}
\end{table}
Then the mass associated with that particular surface Dirac equation can be expressed as
\begin{equation}
m={{({\vec v}_x\times{\vec v}_y)\cdot ({\vec g_x}M_x+{\vec g_y}M_y+{\vec g_z}M_z)}
\over{|{\vec v}_x\times{\vec v}_y|}}.
\end{equation}
Bearing in mind the fact that the magnetization lies in $ab$ plane and using velocity $v$ and g-factor $g$ listed in Table.~\ref{vgtable},
we can express $m$ as a function of magnetization direction
$\theta$ as $m_{010}=-0.3140|\boldsymbol{M}|sin(\theta+\pi/3)$
and $m_{100}=-0.3410|\boldsymbol{M}|sin(\theta-\pi/3)$, which are plotted in Fig.~\ref{schematic}(h).

The mass term vanishes at $\theta=\pi/3$ for the (100) surface and $\theta=2\pi/3$
for (010) indicating the surface topological transitions at these two angles, after which the hinge states
moved from one hinge to another. As we discussed above, the chiral hinge states can only exist
when the two nearby surfaces have the mass terms with opposite signs as indicated by the green area in Fig.~\ref{schematic}(h).

In the first row of Fig.~\ref{hinge_spectra_distr}, we plot the
hinge spectral functions calculated by the recursive Green's function method for four typical magnetization direction, i.e. $0$, $\pi/6$, $\pi/3$ and $\pi/2$.
In Fig.~\ref{hinge_spectra_distr}(a), the magnetization is along the $Y$ direction, which
preserves the $XZ$ mirror plane. As we discussed above, the mirror symmetry guarantees the sign change for the mass terms on the nearby surfaces
leading to chiral hinge states. The energy dispersion of the hinge states on both left and right hinges can be found by checking the spectral functions
 projected to the whole interface area, which are plotted in Fig.~\ref{hinge_spectra_distr}(a).
Clear chiral hinge states can be found with a similar velocity around $-1.60\times10^5 m/s$.
As shown in Fig.~\ref{hinge_spectra_distr}(e) and Fig.~\ref{hinge_spectra_distr}(i), the spacial distribution functions for the
peaks marked as $p_1$ and $p_5$ in Fig.~\ref{hinge_spectra_distr}(a) are centered
at the left and right hinges respectively, which are fully symmetric around the corresponding hinge due to the mirror symmetry.
 Further examination of the spatial distribution of the spectral weight confirms that
the chiral hinge mode on each particular hinge smoothly connects the valence bands on the (100) surface and the conduction bands on the
(010) surface.
When the magnetization angle $\theta$ being rotated away clockwise from zero,
the surface gap on (100) is getting smaller quickly and the hinge states are still there with slightly modified velocity and
asymmetric spacial distribution around the hinge, as shown in Fig.~\ref{hinge_spectra_distr}(f). When the magnetization angle $\theta$ becomes $\pi\over 3$,
the gap on (100) surface closes completely forming surface Dirac cones on the corresponding surfaces, as pointed by the
two arrows in Fig.~\ref{hinge_spectra_distr}(c), which indicates the topological phase transition on these surfaces.
After the transition the connection pattern of the
hinge states changed completely, as shown in Fig.~\ref{hinge_spectra_distr}(d), the hinge states
now connect within the conduction (label $p_4$) or the valence  bands (label $p_8$) in (100) and (010) surfaces and become
topologically trivial.

The existence of chiral hinge modes will cause quantum anomalous Hall effect on the surfaces of the $\text{Sm}_x\text{Bi}_{2-x}\text{Se}_3$. The experimental setup can be
schematically plotted in Fig.~\ref{schematic}(f). By cutting the $\text{Sm}_x\text{Bi}_{2-x}\text{Se}_3$ single crystal using the focused ion beam
 technique to obtain the zig-zag surface structure
formed by (100) and (010) surfaces,
the counter propagating chiral hinge modes can then be induced by the horizontal magnetization on the "ridge" and "valley" area
as shown in Fig.~\ref{schematic}(g). The surface Quantum anomalous Hall effect can be detected by standard four-lead measurement for the Hall effect.

In conclusion,  we proposed in the paper that Sm doped Bi$_2$Se$_3$ single crystal is a magnetic axion insulator and
provides an ideal material platform to realize helical hinge states
on the particular hinges of the crystals. Such nontrivial hinge states demonstrate that the magnetic axion insulator can
also be viewed as higher-order topological insulators with the new type of bulk-boundary correspondence described in the paper. The surface QAHE is the most striking observable effect caused by the chiral hinge states in Sm doped Bi$_2$Se$_3$ single crystal.


\newpage
\begin{widetext}
\appendix
\numberwithin{equation}{section}
\numberwithin{figure}{section}

\section{Calculation of effective magnetic exchange field}\label{sec:magcalc}

Recently, Sm-doped $\text{Bi}_{2}\text{Se}_{3}$ magnetic topological
insulators have been experimentally realized with the Curie temperature
being about 52K \cite{Sm_dopeBi2Se3}. We have simulated this system theoretically by the
first principle method using VASP package \cite{vasp_ref1,vasp_ref2,vasp_ref3}. In the calculation,
a 2$\times$2$\times$1 supercell structure of $\text{Bi}_2\text{Se}_3$ is constructed with one Bi atom being replaced by a Sm atom.
In such a $\text{SmBi}_{23}\text{Se}_{36}$
system, $\text{Sm}^{3+}$ is in the high-spin state and its total
magnetization strength is as large as 5.4 $\mu_{B}$ . A very dense
momentum and energy grid is adopted to calculate the spin polarized density of
states of Bi and Se which are shown in Fig.~\ref{fig_pfdos}. The exchange field
splitting $\Delta E$ of the $p$ orbitals can be roughly estimated
by Eq.(\ref{eq:mag_bydos}), which are about 20 meV and 10 meV for p-orbitals on Se and Bi respectively.
\begin{equation}\label{eq:mag_bydos}
\Delta E=\frac{\int\epsilon\rho_{\downarrow}(\epsilon)d\epsilon}{\int\rho_{\downarrow}(\epsilon)d\epsilon}-\frac{\int\epsilon\rho_{\uparrow}(\epsilon)d\epsilon}{\int\rho_{\uparrow}(\epsilon)d\epsilon},
\end{equation}
where $\rho_{\downarrow(\uparrow)}(\epsilon)$ is the density of state of the majority(minority) spin component of p orbitals.
\begin{figure}[htp]
\includegraphics[clip,width=3.4in,angle=0]{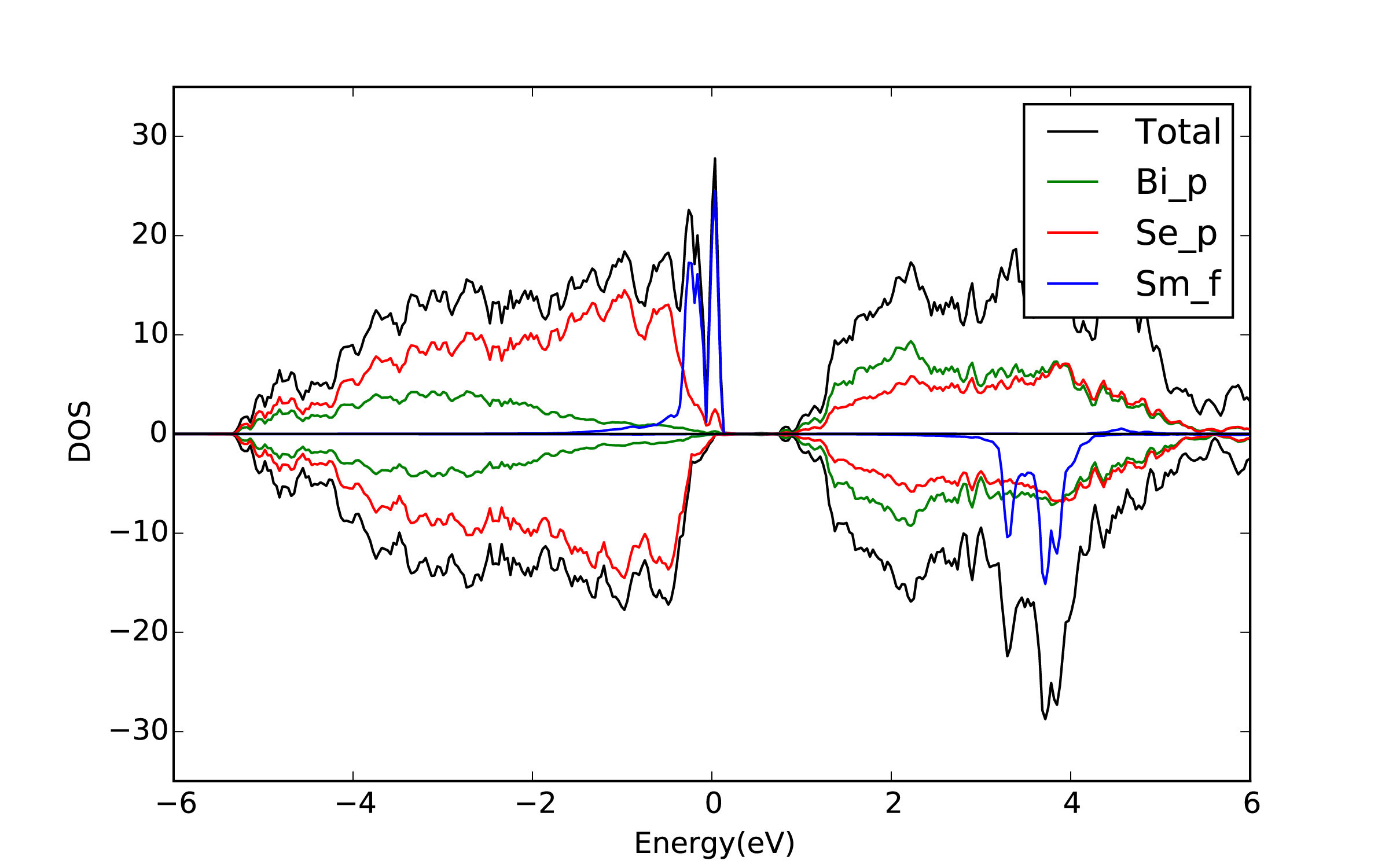}
\caption{Orbital-resolved density of states of $\text{SmBi}_{23}\text{Se}_{36}$ calculated by LDA method.}\label{fig_pfdos}
\end{figure}

\section{Recursive Green Function method in calculation of hinge states}
To carry out the realistic calculations for the hinge states, we design a bi-semi-infite open boundary
geometry in which the (010) and (100) surfaces
meet at \emph{left} and \emph{right} hinges parallel to $\boldsymbol{c}$ direction. The geometry
is semi-infinite along $\boldsymbol{a}$ and $\boldsymbol{b}$ direction, periodic along $\boldsymbol{c}$
direction, finite along $x$ direction. The size along the $x$ direction is $20$ unit cells in realistic
calculation ensuring negligible finite-size effects arising from hybridizations between
left and right hinges. Principal layer (PL) is a group of
atomic layers that is large enough such that only adjacent PLs interact.
Division of ULs and LLs into principal layers is a common strategy to express the Hamiltonian into the block tridiagonal form,

\begin{figure}[htp]
\includegraphics[clip,width=3.4in,angle=0]{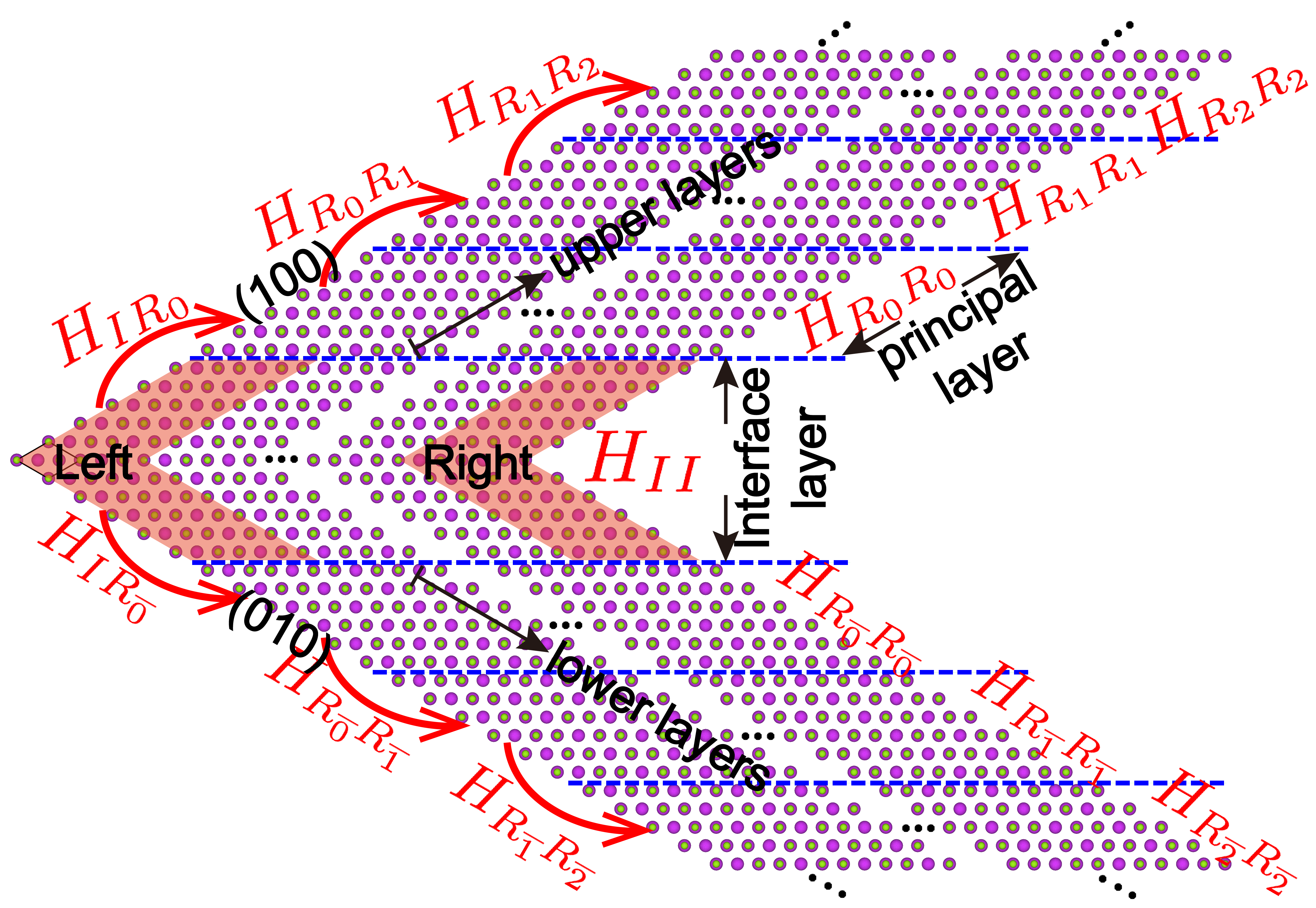}
\caption{
The bi-semi-infinite open boundary geometry mentioned in the main text to calculate hinge states.
The geometry is divided into three parts, \textit{interface layer},
\textit{upper layers} (ULs, divided into PLs which are positively labelled as $R_{i}$) and \textit{lower
layers} (LLs, divided into PLs which are negatively labelled as $R_{\overline{i}}$).
Hamiltonian of each PL and interface layer as well as hopping matrix among them are given in this figure.
}\label{recur_green}
\end{figure}

\begin{equation}\label{eq:Ham_hetero}
H=\left(\begin{array}{cccccccccc}
H_{II} & H_{IR_{0}} & H_{IR_{\overline{0}}} & 0 & \cdots\\
H_{IR_{0}}^{\dagger} & H_{R_{0}R_{0}} & 0 & H_{R_{0}R_{1}} & 0 & \cdots\\
H_{IR_{\overline{0}}}^{\dagger} & 0 & H_{R_{\overline{0}}R_{\overline{0}}} & 0 & H_{R_{\overline{0}}R_{\overline{1}}} & 0 & \cdots\\
0 & H_{R_{0}R_{1}}^{\dagger} & 0 & H_{R_{1}R_{1}} & 0 & H_{R_{1}R_{2}} & 0 & \cdots\\
\vdots & 0 & H_{R_{\overline{0}}R_{\overline{1}}}^{\dagger} & 0 & H_{R_{\overline{1}}R_{\overline{1}}} & 0 & H_{R_{\overline{1}}R_{\overline{2}}} & 0 & \cdots\\
 & \vdots & 0 & H_{R_{1}R_{2}}^{\dagger} & 0 & H_{R_{2}R_{2}} & 0 & H_{R_{2}R_{3}} & 0 & \cdots\\
 &  & \vdots & 0 & H_{R_{\overline{1}}R_{\overline{2}}}^{\dagger} & 0 & H_{R_{\overline{2}}R_{\overline{2}}} & 0 & H_{R_{\overline{2}}R_{\overline{3}}} & \ddots\\
 &  &  & \vdots & \ddots & \ddots & \ddots & \ddots & \ddots & \ddots\\
\\
\\
\end{array}\right)
\end{equation}
where the diagonal block $H_{R_{i}R_{i}}=H_{R_{0}R_{0}}$, $H_{R_{\overline{i}}R_{\overline{i}}}=H_{R_{\overline{0}}R_{\overline{0}}}$
and the hopping matrix $H_{R_{i}R_{i+1}}=H_{R_{0}R_{1}}$,
$H_{R_{\overline{i}}R_{\overline{i+1}}}=H_{R_{\overline{0}}R_{\overline{1}}}$
Implicitly, each block of $H$ is a function of momentum $k_{z}$,
i.e. $H_{\cdot\cdot}\equiv H_{\cdot\cdot}(k_{z})$.
The tight binding Hamiltonian is obtained from the maximally-localized Wannier functions constructed by wannier90 \cite{wannier90}
package interfaced to VASP \cite{vasp_ref1,vasp_ref2,vasp_ref3}.
Furthermore, Zeeman terms $H_z=\boldsymbol{M}\cdot\boldsymbol{\sigma}$ with the strength obtained from the first principle calculations mentioned in the
 previous section are added to all $p$
orbitals to simulate the ferromagnetic order arising from Sm doping.

For simplification, $H_{IR}$ is used to denote hoppings between interface
layer and ULs, LLs
\begin{equation}
H_{IR}\equiv\left(\begin{array}{cccc}
H_{IR_{0}} & H_{IR_{\overline{0}}} & 0 & \cdots\end{array}\right).
\end{equation}
$H_{P}$, in block tridiagonal form, is to denote Hamiltonian
of ULs,
\begin{equation}
H_{P}\equiv\left(\begin{array}{ccccc}
H_{R_{0}R_{0}} & H_{R_{0}R_{1}} & 0\\
H_{R_{0}R_{0}}^{\dagger} & H_{R_{0}R_{0}} & H_{R_{0}R_{1}} & 0\\
0 & H_{R_{0}R_{0}}^{\dagger} & H_{R_{0}R_{0}} & H_{R_{0}R_{1}} & \ddots\\
 & \ddots & \ddots & \ddots & \ddots\\
\\
\end{array}\right),
\end{equation}
and $H_{N}$ , also in block tridiagonal form, is to denote
Hamiltonian of LLs
\begin{equation}
H_{N}\equiv\left(\begin{array}{ccccc}
H_{R_{\overline{0}}R_{\overline{0}}} & H_{R_{\overline{0}}R_{\overline{1}}} & 0\\
H_{R_{\overline{0}}R_{\overline{1}}}^{\dagger} & H_{R_{\overline{0}}R_{\overline{0}}} & H_{R_{\overline{0}}R_{\overline{1}}} & 0\\
0 & H_{R_{\overline{0}}R_{\overline{1}}}^{\dagger} & H_{R_{\overline{0}}R_{\overline{0}}} & H_{R_{\overline{0}}R_{\overline{1}}} & \ddots\\
 & \ddots & \ddots & \ddots & \ddots\\
\\
\end{array}\right).
\end{equation}
Furthermore, we introduce a new ``direct sum'' operation $\tilde{\oplus}$
of two square matrix $A$ and $B$ as
\begin{equation}
A\tilde{\oplus}B=\left(\begin{array}{ccccc}
A_{11} & 0 & A_{12} & 0 & \cdots\\
0 & B_{11} & 0 & B_{12} & \cdots\\
A_{21} & 0 & A_{22} & 0 & \cdots\\
0 & B_{21} & 0 & B_{22} & \cdots\\
\vdots & \vdots & \vdots & \vdots & \ddots
\end{array}\right).
\end{equation}
Now we have

\begin{equation}
H\equiv\left(\begin{array}{cc}
H_{II} & H_{IR}\\
H_{IR}^{\dagger} & H_{RR}
\end{array}\right),
\end{equation}
where
\begin{equation}
H_{RR}\equiv H_{P}\tilde{\oplus}H_{N},
\end{equation}
indicating that ULs and LLs are totally decoupled from each other.

The imaginary part of the
interface Green function $G_{II}(k_{z},\omega)$ can be written as,

\begin{equation}\label{eq:AII}
A_{II}(k_{z},\omega)=-\frac{1}{\pi}\Im G_{II}(k_{z},\omega),
\end{equation}
\begin{equation}
G\equiv\left(\begin{array}{cc}
G_{II} & G_{IR}\\
G_{IR}^{\dagger} & G_{RR}
\end{array}\right),
\end{equation}
\begin{equation}
(\omega+i\eta)G=1,
\end{equation}
\begin{equation}\label{eq:GII}
G_{II}(k_{z},\omega+i\eta)=\left[(\omega+i\eta)-H_{II}-\Sigma_{R}\right]^{-1},
\end{equation}
where $\Sigma_{R}$ is the self-energy depicting comprehensive interactions
between interface layer and ULs, LLs. Within the approximation of
PL, $\Sigma_{R}$ has a simple form as
\begin{equation}\label{eq:SigmaRR}
\Sigma_{R}=H_{IR}g_{RR}H_{IR}^{\dagger}\approx H_{IR_{0}}g_{R_{0}R_{0}}H_{IR_{0}}^{\dagger}+H_{IR_{\overline{0}}}g_{R_{\overline{0}}R_{\overline{0}}}H_{IR_{\overline{0}}}^{\dagger}
\end{equation}
where $g_{R_{0}R_{0}}$ and $g_{R_{\overline{0}}R_{\overline{0}}}$
is the ``surface'' Green function of ULs and LLs, respectively.
Because $H_{P}$ and $H_{N}$ are all in triangular diagonal form,
$g_{R_{0}R_{0}}$ and $g_{R_{\overline{0}}R_{\overline{0}}}$ can
be solved by the standard recursive schemes \cite{Green_recur1985}.
Using equations (\ref{eq:SigmaRR}, \ref{eq:GII}, \ref{eq:AII}), we obtain the spectral functions for the
interface layer.  $A_{II}$ is a square matrix with its indices being the number of orbitals
in the interface layer. The trace of $A_{II}$ gives the integrated spectral shown in the first row of 
Fig.~\ref{hinge_spectra_distr}. 

\end{widetext}

\clearpage
\bibliography{bs_ref.bib}

\end{document}